\documentclass[aps,prl,reprint]{revtex4-2}
\pdfoutput=1

\usepackage[T1]{fontenc}
\usepackage{graphicx,amsmath,amssymb,amsthm,mathtools,xcolor,hyperref,bookmark,makecell}
\usepackage[caption=false]{subfig}

\definecolor{dark-blue}{rgb}{0.3,0.3,0.7}
\definecolor{dark-red}{rgb}{0.7,0,0}
\hypersetup{
    colorlinks=true,
    linkcolor=dark-blue,
    citecolor=dark-red,
    urlcolor=dark-blue,
    linktoc=page
}

\renewcommand*{\toprule}{\Xhline{0.075em}\noalign{\vskip 0.65ex}}
\newcommand*{\midrule}{\noalign{\vskip 0.525ex}\Xhline{0.045em}\noalign{\vskip 0.65ex}}
\newcommand*{\bottomrule}{\noalign{\vskip 0.525ex}\Xhline{0.075em}}

\newcommand*{\di}{\mathrm{d}}
\newcommand{\abs}[1]{\mathopen{|}#1\mathclose{|}}

\begin{document}
\setlength{\abovedisplayshortskip}{5pt plus 2pt minus 2pt}

\title{The Kontsevich--Segal Criterion in the No-Boundary State Constrains Inflation}

\author{Thomas Hertog$^\spadesuit$, Oliver Janssen$^\clubsuit$ and Joel Karlsson$^\spadesuit$\vspace{0.1cm}}

\affiliation{$^\spadesuit${\small\slshape Institute for Theoretical Physics, KU Leuven, Celestijnenlaan 200D, 3001 Leuven, Belgium} \\
$^\clubsuit${\small\slshape International Centre for Theoretical Physics, Strada Costiera 11, 34151 Trieste, Italy} and \\
{\small\slshape Institute for Fundamental Physics of the Universe, Via Beirut 2, 34014 Trieste, Italy}}

\begin{abstract} \noindent
We show that the Kontsevich--Segal (KS) criterion, applied to the complex saddles that specify the semiclassical no-boundary wave function, acts as a selection mechanism on inflationary scalar field potentials. Completing the observable phase of slow-roll inflation with a no-boundary origin, the KS criterion effectively bounds the tensor-to-scalar ratio of cosmic microwave background fluctuations to be less than 0.08, in line with current observations. We trace the failure of complex saddles to meet the KS criterion to the development of a tachyon in their spectrum of perturbations.
\end{abstract}

\maketitle

\noindent {\it Dedicated to the memory of Jim Hartle, whose innate quantum outlook on cosmology will be a source of inspiration for many years to come.}

\section{Introduction} \label{introsec} \noindent
Future experiments promise to tighten the upper bound on the tensor-to-scalar ratio $r$ of CMB fluctuations down to around $10^{-3}$ \cite{CMB2022,abazajian2016cmbs4}. It is therefore of great experimental and theoretical interest to understand whether quantum gravity can produce inflationary models with a level of primordial gravitational waves above this.

The value of $r$ predicted by inflationary theory is related to the total amount of displacement experienced by the inflaton field as it rolls down the scalar potential. While the first string theory models of inflation predicted an unobservably low tensor contribution to the CMB anisotropies, further studies have suggested that $r$ might lie in the detectable range after all \cite{Silverstein:2016ggb,Baumanncosmo}. Still, many suspect that the quantum completion of inflation implies a theoretical upper bound on $r$ and it would be very interesting to understand where it is.

Recently, Kontsevich and Segal (KS) in \cite{Kontsevich:2021dmb} have advanced an interesting criterion that complex metrics should satisfy in order to qualify as backgrounds for physically meaningful quantum field theories. Witten \cite{Witten:2021nzp} subsequently explored whether this criterion might be employed to select physically sensible saddles of gravitational path integrals (see \cite{Halliwell:1989dy,Louko:1995jw} for earlier work in this direction). The idea behind this is that only those backgrounds $g$ on a $D$-manifold $M$ should be considered (or are \mbox{``allowable''}) on which an arbitrary quantum field theory could be defined. Concretely one takes this to mean that the path integral for all free $p$-form matter on $(M,g)$ should converge, or that
\begin{align} \label{allowable1}
    &\operatorname{Re} \left( \sqrt{g} \, g^{\mu_1 \nu_1} \cdots g^{\mu_p \nu_p}  F_{\mu_1 \cdots \mu_p} F_{\nu_1 \cdots \nu_p} \right) > 0
\end{align}
for all $p \in \{0,\cdots,D\}$ and all \textit{real-valued} antisymmetric $p$-tensors $F$ on $M$. For metrics that are diagonal in a real basis \footnote{Metrics that satisfy the criterion can always be diagonalized in a real basis, but the metrics we will consider in the following are already in diagonal form from the get-go. Notice that the criterion \eqref{allowable1} should hold pointwise on $M$. Further, the criteria arising from $p = q$ and $p = (D-q)$-forms are equivalent: the former has $q$ minus signs in the sum $\abs{\pm \arg \lambda_1 \pm \arg \lambda_2 \pm \cdots \pm \arg \lambda_D} < \pi$ while the latter has $D-q$. Finally, the $p = 0$ criterion reads $\operatorname{Re} \sqrt{g} > 0$ and can be motivated from the convergence of the path integral for a massive scalar on $(M,g)$.} with diagonal elements $\lambda_i$, this is equivalent \cite{Kontsevich:2021dmb} to the requirement that
\begin{equation} \label{allowable2}
    \sum_{i=1}^D \left| \arg \lambda_i \right| < \pi \,,
\end{equation}
where $\arg \in (-\pi,\pi]$.

\begin{figure}[t!]
\centering
\includegraphics[width=86.45mm]{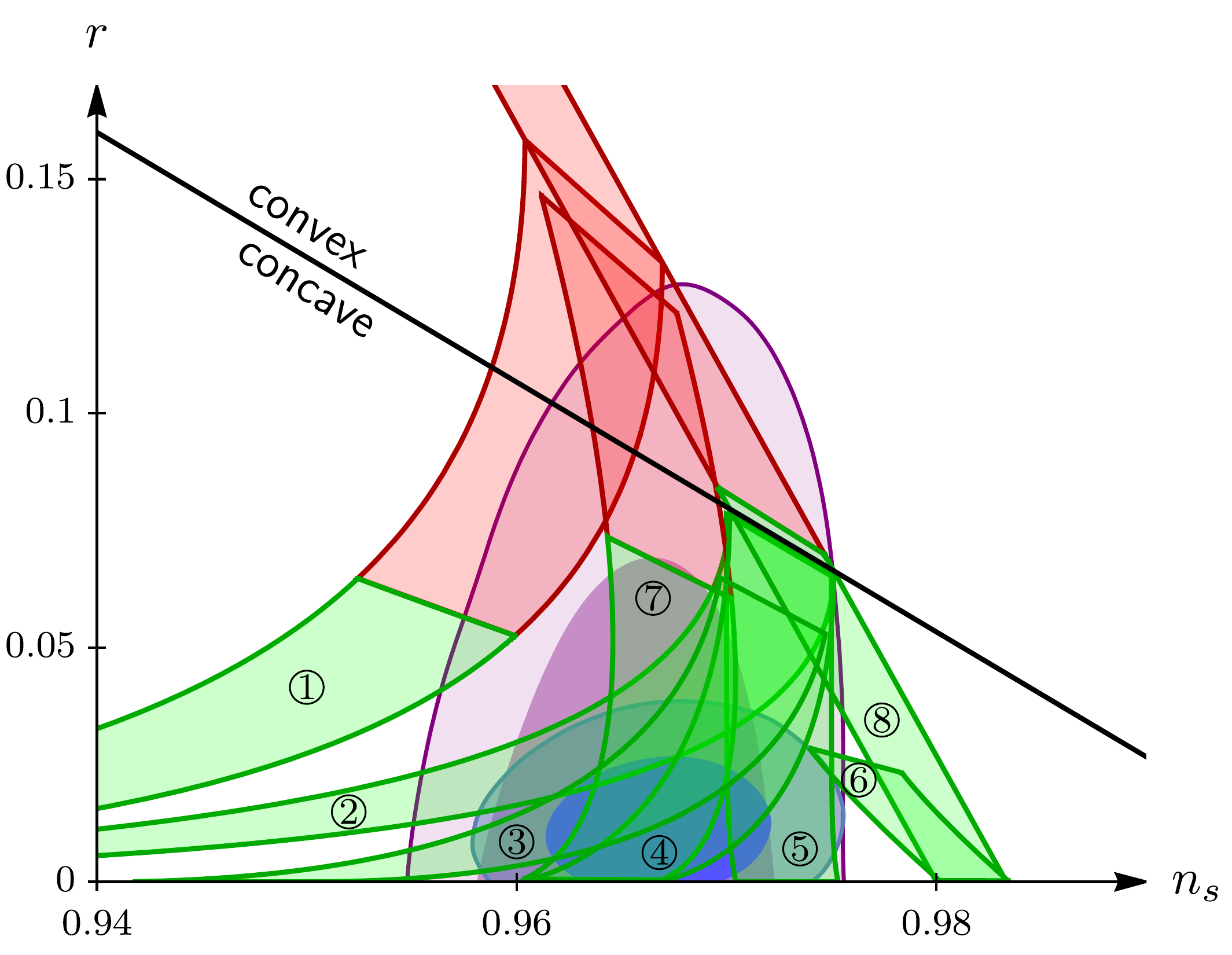}
\caption{The Kontsevich--Segal criterion applied to the no-boundary state selects those inflationary models that predict CMB fluctuations with a low tensor-to-scalar ratio $r \lesssim 0.08$. Shown are the predictions for the scalar tilt $n_s$ and $r$ in eight different slow-roll models of inflation (see Table~\ref{thetable}). Indicated in green are inflationary trajectories of 50 to 60 $e$-folds that are associated with no-boundary saddles that satisfy the KS criterion. Shown in red are inflationary universes that are ruled out by KS. The observational constraints from the 2018 Planck TT,TE,EE+lowE+lensing analysis are indicated in purple. Finally, the blue region shows the combined constraints of Planck and the 2018 BICEP/Keck data and BAO.}
\label{thefigure}
\end{figure}

This criterion has passed several nontrivial checks \cite{Witten:2021nzp}. For example, it eliminates pathological wormhole solutions with vanishing action, but it does allow for the complexified Kerr solutions that correctly encode the thermodynamic properties of rotating black holes. Yet it remains unclear whether the KS criterion is necessary or sufficient. Regarding necessity, recently solutions were found \cite{Maldacena:2019cbz,Bah:2022uyz} which violate the KS criterion but nonetheless appear to describe physically sensible saddles. Regarding sufficiency, clearly not all sensible quantum field theories are covered by those of free $p$-forms. Hence more work is needed both to refine the KS criterion and to better understand its physical implications, especially in the context of inflation (for recent studies see \cite{Lehners:2022xds,Jonas:2022uqb,Lehners:2021mah,Visser:2021ucg,Briscese:2022evf}).

To this end we study the implications of the KS criterion for our understanding of the quantum gravitational origin of inflation. We assume the universe to be in the Hartle--Hawking no-boundary state \cite{Hartle:1983ai} and we consider this wave function in a variety of single field, slow-roll models of inflation. In each of these models the semiclassical no-boundary wave function (NBWF) is specified by $O(4)$-invariant, complex solutions of the Einstein equations. Loosely speaking, these complex saddles describe the nucleation and subsequent quasiclassical evolution of an expanding universe with an early phase of inflation.

However, while the original NBWF implies an inflationary origin, the theory allows for a vast range of inflationary potentials. That is, even though the no-boundary prior favors some potentials over others (e.g.~\cite{Hertog2014}), it does not exclude slow-roll potentials entirely. We find that the KS criterion does exactly this. In the next sections, we show that the KS criterion in conjunction with no-boundary initial conditions acts as a selection mechanism on inflationary potentials. Specifically, the criterion predicts that universes with a significant number of $e$-folds emerge from a concave patch of the potentials. This in turn sets an upper bound on the tensor-to-scalar ratio of CMB fluctuations, consistent with observations. Figure~\ref{thefigure} summarizes our findings. We now describe how we arrive at these.

\section{Allowable no-boundary saddles} \label{NBsec} \noindent
We consider the Hartle--Hawking state in $D = 4$ in a minisuperspace model consisting of Einstein gravity minimally coupled to a homogeneous scalar field with potential $V$. When it exists, we assume that the semiclassical no-boundary amplitude $\Psi(b,\chi)$ of a round three-sphere with volume $\propto b^3$ filled with a uniform scalar field of value $\chi$ is specified by an $O(4)$-symmetric saddle living on the four-ball $M = B^4$ and satisfying the KS criterion. That is, we follow \cite{Halliwell:2018ejl} but include KS.

We adopt the following Ansatz for the saddle-point geometries and field profile,
\begin{equation}
    g_{\mu \nu}\, \di x^\mu\, \di x^\nu = \di r^2 + a(r)^2\, \di \Omega_3^2 \,, \quad
    \phi = \phi(r) \,, \label{NBbackground}
\end{equation}
where the scale factor $a$ and scalar field $\phi$ take the values $(b,\chi)$ on the boundary $\partial M = S^3$.
The saddle-point equations of motion (EOM) are (with $M_\mathsf{pl} = 1$)
\begin{align}
    \left( \frac{a'}{a} \right)^2 &= \frac{1}{a^2} + \frac{1}{3} \left( \frac{(\phi')^2}{2} - V(\phi) \right) , \notag \\
    0 &= \phi'' + 3 \frac{a'}{a} \phi' - V'(\phi) \,. \label{EOM}
\end{align}
The coordinate $r$ runs from the center of the ball at $r = 0$ to an endpoint at $r = v$. Its range is determined by the boundary conditions of regularity at the center,
\begin{equation}
    a(r) = r + \mathcal{O}(r^3) \,, \;\;
    \phi(r) = \phi_0 + \mathcal{O}(r^2) \;\;\;
    \text{as } r \rightarrow 0 \,, \label{initSP}
\end{equation}
together with the conditions that
\begin{equation}
    a(v) = b \,,\quad \phi(v) = \chi \,. \label{finalBD}
\end{equation}
These boundary conditions generally imply that $\phi_0$ and $v$ are complex \cite{Hartle:2008ng,Janssen:2020pii} and hence that the solutions $[a(r),\phi(r)]$ are complex too. Note that this need not be at odds with the assumption underlying the KS criterion that one integrates over real matter field fluctuations, since complex saddle-point solutions can arise as an approximation to an integral over real $\{ \phi \}$.

The $r$-coordinate runs along a curve $\gamma(\ell) : 0 \rightarrow v$ in the complex plane. Along this curve the metric reads
\begin{equation}
    \di s^2 = \gamma'(\ell)^2\, \di \ell^2 + a(\gamma(\ell))^2\, \di \Omega_3^2 \,. \label{induced}
\end{equation}
We say that a given solution $[a(r),\phi(r),v]$ obeys the KS criterion if there exists a curve $\gamma$ such that the induced metric \eqref{induced} satisfies \eqref{allowable2} along its entire length:
\begin{equation} \label{allowable3}
    \abs{ \arg \gamma'^2} + 3 \abs{ \arg a(\gamma)^2} < \pi \,.
\end{equation}

The boundary value problem \eqref{EOM}--\eqref{finalBD} has two complex boundary conditions in \eqref{finalBD} and equally many free parameters in $(\phi_0,v)$. Hence it has a discrete solution set. Furthermore for given $(b,\chi)$, each solution $[a(r),\phi(r),v]$ is fourfold degenerate: the tuples $[a(-r),\phi(-r),-v]$, $[a(r^*)^*,\phi(r^*)^*,v^*]$ and $[a(-r^*)^*,\phi(-r^*)^*,-v^*]$ are also solutions. Either they all satisfy the KS criterion or none of them do, as it should be because the physical predictions of the four saddles are identical (see \S\ref{discussionsec}). In all the models we will consider (see \S\ref{examplessec}-\ref{SRsec}), we find an $O(4)$-symmetric solution to \eqref{EOM}--\eqref{finalBD} for all $(b,\chi)$, but that solution does not necessarily satisfy KS.

Considering the solution with $v$ in the first quadrant, our strategy to verify the KS criterion is based on the construction of an ``extremal curve'' $\gamma_\mathsf{e}$ that saturates the inequality \eqref{allowable3} and lies in the first quadrant (cf.~\cite{Jonas:2022uqb}):
\begin{equation} \label{extremalcurve}
    \arg \gamma_\mathsf{e}'^2 + 3 \abs{ \arg a(\gamma_\mathsf{e})^2 } = \pi \,.
\end{equation}
From the known behavior \eqref{initSP} of the scale factor near the origin it follows that $\lim_{\ell \rightarrow 0} \arg \gamma_\mathsf{e}(\ell) = \pi/8$. Note that $\gamma_\mathsf{e}$ is required to always be right-moving. Also, the curves everywhere satisfying \eqref{allowable3} and starting at $r = 0$ are constrained to remain below $\gamma_\mathsf{e}$. Therefore if $\operatorname{Im} \gamma_\mathsf{e} < \operatorname{Im} v$ when $\operatorname{Re} \gamma_\mathsf{e} = \operatorname{Re} v$, there is no allowable $\gamma : 0 \rightarrow v$. Conversely if $\operatorname{Im} \gamma_\mathsf{e} \geq \operatorname{Im} v$ when $\operatorname{Re} \gamma_\mathsf{e} = \operatorname{Re} v$, we expect by continuity there to exist an allowable curve $\gamma : 0 \rightarrow v$, obtained from \eqref{extremalcurve} by decreasing the right-hand side.

For any given $(b,\chi)$ this procedure allows us to determine whether there is an $O(4)$-invariant no-boundary saddle that meets the KS criterion \cite{supplemental}. A systematic analysis for all $(b,\chi)$ thus divides the minisuperspace in two regions. One range of configurations will be associated with saddles that are physically meaningful, according to the KS criterion. The semiclassical amplitude of these configurations is specified by the usual Hartle--Hawking saddle. But in regions of superspace where the KS criterion fails, the original ``vanilla'' NBWF will be strongly modified. Specifically, the KS criterion strongly suppresses the semiclassical NBWF in this regime, by excluding the contribution from what would have been the dominant saddle in the absence of the KS criterion. This in turn sharpens the predictions of the theory \footnote{The semiclassical, KS-corrected NBWF need not vanish in the regime where KS excludes the leading saddle, since there may be another instanton with less symmetry and a larger action, satisfying KS, that contributes to the wave function in this regime. We leave an exploration of this scenario for future work.}.

We now carry out the above analysis, first in a particular model that is analytically solvable and then in a representative class of slow-roll inflation models.

\section{A solvable model} \label{examplessec} \noindent
Consider Einstein gravity minimally coupled to a scalar subject to the potential
\begin{equation}
    V(\phi) = \Lambda \cosh\Biggl( \sqrt{\frac{2}{3}} \phi \Biggr) , \quad \Lambda > 0 \,.
\end{equation}
Note that this potential does not have standard slow-roll patches, since $\eta = V''/V = 2/3$ everywhere. Rather it is the combination of the cosmological constant and the scalar field in the lower regions of this potential that can drive exponential expansion.

A change of coordinates $\di r \rightarrow \di \tau/a$, together with an overall rescaling so that
\begin{equation} \label{newmetric}
    \di s^2 = \frac{\sqrt{3/2}}{\Lambda} \left( \frac{\di \tau^2}{a(\tau)^2} + a(\tau)^2 \di \Omega_3^2 \right) ,
\end{equation}
and the introduction of new variables
\begin{equation} \label{xycoords}
    x = \sqrt{\frac{3}{2}} \, a^2 \cosh \Biggl( \sqrt{\frac{2}{3}} \phi \Biggr) , \quad y = \sqrt{\frac{3}{2}} \, a^2 \sinh \Biggl( \sqrt{\frac{2}{3}} \phi \Biggr)
\end{equation}
results in a quadratic Euclidean action for $(x,y)$ \cite{Garay:1990re,DiazDorronsoro:2017hti}:
\begin{equation}
    S = \frac{\sqrt{6}\, \pi^2 }{\Lambda} \int \di \tau \left[ \frac{1}{2} \left( \dot{y}^2 - \dot{x}^2 \right) + x - 3 \right] .
\end{equation}
The EOM with boundary conditions $x(0) = y(0) = 0$, $x(v) = X$, $y(v) = Y$, where $(X,Y)$ are specified by $(b,\chi)$ through \eqref{xycoords}, are solved by
\begin{equation}
    x(\tau) = -\frac{\tau^2}{2} + A \tau \,, \qquad
    y(\tau) = B \tau \,,
\end{equation}
where
\begin{equation}
    A = \frac{1}{v} \left( X + \frac{v^2}{2} \right) , \quad B = \frac{Y}{v} \,,
\end{equation}
while the Hamiltonian constraint determines the possible values of the endpoint $v$ in the complex $\tau$-plane. The solution of interest in the first quadrant is given by
\begin{equation} \label{veq}
    v = \sqrt{C+D} + \sqrt{C-D}
\end{equation}
where
\begin{equation}
    C = 6-X \,, \quad D = \sqrt{X^2 - Y^2} \,.
\end{equation}
The equation for the extremal curve in these coordinates reads
\footnote{Numerical exploration indicates that dropping the absolute value in the second term is consistent.}
\begin{equation} \label{extremalcurvecosh}
    \arg \gamma_\mathsf{e}'^2 + 2 \arg a(\gamma_\mathsf{e})^2 = \pi \,,
\end{equation}
where, via \eqref{xycoords}, $a = \left[ 2(x^2 - y^2)/3 \right]^{1/4}$.

In the regime $C \geq D$, which essentially corresponds to $b H(\chi) < \mathcal{O}(1)$ where $H(\chi) = \sqrt{V(\chi)/3}$, $v$ in \eqref{veq} is seen to lie on the positive real axis. In this regime the metric \eqref{newmetric} is purely Euclidean on the segment $[0,v]$. Hence the semiclassical amplitude of all such configurations $(b,\chi)$ is given by a no-boundary solution that meets the KS criterion.

In the regime $C \leq -D$ on the other hand, which corresponds to $b H > \mathcal{O}(1)$ and $b \chi > 6^{3/4}$ when $\chi \ll 1$, $v$ is purely imaginary. From \eqref{extremalcurvecosh} it follows there can be no curve that connects the origin to $v$ along which the induced metric satisfies \eqref{allowable2}. Thus the KS criterion, taken at face value, appears to strongly suppress the semiclassical amplitude of this part of the minisuperspace, by excluding what would have been the leading saddle.

Finally we have the intermediate regime $|C| < D$, which corresponds to $b H > \mathcal{O}(1)$ and $b \chi < 6^{3/4}$ when $\chi \ll 1$, and which includes the de Sitter (dS) solution with $\chi= 0$. Here $v$ is neither real nor imaginary but complex. A solution for $\gamma_\mathsf{e}$ in \eqref{extremalcurvecosh} is given by $\gamma_\mathsf{e}' = i/a(\gamma_\mathsf{e})^2$, which upon integration gives the relation
\begin{align}
    &\frac{1}{3} \left( \gamma_\mathsf{e}^2 - A\gamma_\mathsf{e} - 6B^2-12 \right) \sqrt{\gamma_\mathsf{e}^2 - 4 A\gamma_\mathsf{e}+24} \notag \\
    &-4AB^2 \tanh^{-1}\Biggl( \frac{\gamma_\mathsf{e} - 2A}{\sqrt{\gamma_\mathsf{e}^2 - 4 A\gamma_\mathsf{e}+24}} \Biggr) = i \sqrt{6} \, \ell + \mathsf{const.} \,, \label{cosheq}
\end{align}
where the constant is determined by setting $\gamma_\mathsf{e}(0) = 0$. To proceed, we set $\gamma_\mathsf{e} = v$ in \eqref{cosheq} and equate the real parts of both sides. This yields a curve $\chi_\star(b)$ that indicates those points which the KS criterion marginally allows. Points lying above this curve in the $(b,\chi)$-plane are excluded while points below it are allowable. Asymptotically this critical line behaves as
\begin{equation} \label{intermed}
    \chi_\star(b) = \frac{6^{1/4}}{b \sqrt{\log b}} \left[ 1 + \mathcal{O} \biggl( \frac{1}{\log b} \biggr) \right] \quad \text{as } b \rightarrow \infty \,.
\end{equation}

On the other hand one can examine the set of classical histories predicted by $\Psi$ in this model \cite{Halliwell:1989myn,Hartle:2008ng,DiazDorronsoro:2017hti}. These are the curves $p_\alpha = \partial_\alpha \operatorname{Im} S$, where $S$ is the action of the complex saddle and $p_\alpha$, $\alpha \in \{ a,\phi \}$, are the canonical momenta. In a 2D minisuperspace model of this kind, this is a one-parameter set of curves, or histories, in the $(b,\chi)$-plane. In the regime $|C| < D$, these histories are characterized by the relation
\begin{equation}
    \chi_\mathsf{classical}(b) = c \, \frac{6^{3/4}}{b} \bigl(1 + \mathcal{O}(1/b) \bigr) \quad \text{as } b \rightarrow \infty
\end{equation}
with $c \in [0,1)$ labeling the history, where $c = 0$ corresponds to dS space (the dependence on $b$ can be inferred from \eqref{EOM} after rotating $r \rightarrow it$). Comparison with \eqref{intermed} shows that, strikingly, every classical history except empty dS exits the domain of allowability at some point. That is, the no-boundary state augmented by the KS criterion predicts that classical evolution does not continue forever in this model. It would be interesting to better understand whether this is a peculiar property of this particular model or a more general prediction of the KS criterion in conjunction with no-boundary conditions. In a realistic cosmology, however, this would require one to take into account the coupling of the inflaton to other forms of matter in order to evaluate the wave function well after inflation ends -- and indeed at the present stage of evolution.

\section{Slow-roll inflation} \label{SRsec} \noindent
We now turn to the no-boundary saddles that appear in slow-roll models of inflation \cite{Hartle:2008ng,Janssen:2020pii}. We are especially interested in regions of the minisuperspace where the scale factor is large in local Hubble units, $b H(\chi) \gg 1$ with $H(\chi) \approx \sqrt{V(\chi)/3}$ in the slow-roll regime. Based on our results above, we expect that as the potential becomes flatter, or more precisely, as the background $a(r)$ approaches the form
\begin{equation} \label{HHdS}
    a_\mathsf{dS}(r) = \frac{1}{H} \sin(H r)
\end{equation}
with $H$ constant, more $e$-folds $\log b H(\chi)$ will be allowable.

The saddles $[a(r),\phi(r),v]$ that correspond to configurations $(b,\chi)$ along a slow-roll trajectory are complex deformations of the so-called real tunneling instanton \eqref{HHdS} that describes the quantum creation of empty dS. In its familiar representation the dS saddle consists of half of a four-sphere of radius $1/H$, along the segment $[0,\pi/2H]$ of the real $r$-axis, glued to the expanding branch of Lorentzian dS space along a segment parallel to the imaginary $r$-axis \footnote{This particular representation does not obey the KS criterion, but a deformation of the $r$-contour renders this saddle allowable (see \S\ref{examplessec}, and also \cite{Witten:2021nzp}).}. No-boundary saddles associated with slow-roll inflationary universes typically involve half of a deformed $S^4$ with an approximate radius $1/H(\abs{\phi_0})$, which transitions to a slow-roll attractor in the imaginary $r$-direction. Importantly, $a(r)$ and $\phi(r)$ are purely real along neither segment except if the inflaton starts out at an extremum of $V$. Instead, along the approximately Lorentzian direction, the imaginary parts decay in a way dictated by the real parts, whose evolution is governed by the usual slow-roll approximation, viz.\ $\operatorname{Im} a \sim \left( \operatorname{Re} a \right)^{-2}$, $\operatorname{Im} \phi \sim \left( \operatorname{Re} a \right)^{-3}$ \cite{Janssen:2020pii}. It is this mere approximate reality of the fields that can cause the KS criterion to fail for certain configurations $(b,\chi)$  \footnote{That this is consistent follows from the following argument: suppose $a(x+i \ell)$ is real and increasing for all $\ell$ at a fixed $x$. Then $a'(x+i\ell)$ is negative imaginary. To solve for the extremal curve we write $\gamma_\mathsf{e}(\ell) = x - \varepsilon(\ell) + i \ell$ and expand at large $\ell$, assuming $\varepsilon$ is small, and initially positive, so that $a(\gamma(\ell)) \approx a(x+i\ell) - a'(x+i\ell) \varepsilon$. With this one checks that the allowability criterion \eqref{allowable2} becomes $\varepsilon' \approx -3 \left( - \operatorname{Im}(a')/a \right) \varepsilon$, so that indeed $\varepsilon$ is decreasing, perhaps to zero, but does not change sign.}.

We proceed by solving the equations governing the background \eqref{EOM}--\eqref{finalBD} and the extremal curve \eqref{extremalcurve} numerically \cite{supplemental}. A trustworthy analysis of the KS criterion requires exponential numerical precision. This can be seen even from the pure dS saddle, where the extremal curve asymptotes to the vertical line $\operatorname{Re} r = \pi/2H$ on which the endpoints $v(b)$ are located, as
\begin{equation} \label{dSasymptotics}
    \operatorname{Re}\biggl( \frac{\pi}{2H} - \gamma_\mathsf{e} \biggr) = \mathcal{O} \biggl( \frac{1}{H} \exp \left( -3H \, \operatorname{Im} \gamma_\mathsf{e} \right) \biggr) ~ \text{as } \operatorname{Im} \gamma_\mathsf{e} \rightarrow \infty \,.
\end{equation}
Interestingly, this is the sort of level of detail through which no-boundary saddle-point geometries in the large-volume regime encode the fine details of the quantum origin of inflation. Hence, physically the required accuracy stems from the fact that KS is a global criterion on complex saddles that probes the quantum nature of inflation, even at late times.

We would like to determine in which models saddles corresponding to inflationary histories with $N_e = \mathcal{O}(50-60)$ $e$-folds meet the KS criterion. To identify these models we first pick a potential and fix $\chi$ to its value at the end of inflation, where $\varepsilon = (V'/V)^2/2$ or $|\eta| = |V''|/V$ are equal to unity. Then we vary $\log b H$ between 50 and 60. Finally we use the method described in \S\ref{NBsec} to verify whether the no-boundary saddle corresponding to these configurations $(b,\chi)$ is allowable.

We carried out this procedure for most of the inflationary potentials discussed in the 2018 Planck analysis \cite{Planck:2018jri}. We ensured all the numerics are trustworthy by dialing up the precision of our numerical algorithm \cite{supplemental} and observing convergence in the results. Note that the KS criterion does not depend on the overall scale of the potentials \cite{supplemental}, which may thus be adjusted to match the observed amplitude of CMB fluctuations.

We summarize some of our results in Table~\ref{thetable}, where for eight one-parameter potentials we list the ranges of parameter values $f, \mu, \dots$ for which the KS criterion applied to saddles with $N_e = 60$ $e$-folds is satisfied. As an example, consider the power-law potentials $V \propto \phi^p$. Whereas the KS criterion allows inflationary histories with $N_e = 60$ for $p \lesssim 1.05$, for larger values of $p$ we find that all slow-roll saddles exit the regime of allowability before the end of inflation \footnote{To verify this last claim requires a different numerical setup, namely setting $(b,\chi)$ to its values along a slow-roll history. No-boundary instantons which prepare small universes $(b,\chi) \approx (1/H(\chi_0),\chi_0)$ satisfy KS because they are essentially Euclidean. Further along the slow-roll trajectory, KS may fail.}. In general, the table shows that the KS criterion selects those universes in the no-boundary state that emerge on a \textit{concave} patch of the scalar slow-roll potential, with an additional model-dependent pressure towards lower values of $r$. This in turn favors small-field models of inflation \cite{Lyth:1996im}.

\begin{table}[t!]
\centering
\caption{Families of slow-roll potentials in which we subjected the no-boundary instantons giving rise to 60 $e$-folds of inflation to the KS criterion. In the ``allowable'' column we list the ranges of parameter values $f, \mu, \dots$ that specify potentials in which configurations $(b,\chi)$ at the end of 60 $e$-folds of inflation, prepared by no-boundary conditions, satisfy the KS criterion.  The ``disallowable'' column lists parameter values for which the criterion is not satisfied.}
\begin{tabular}{clll}
    \toprule
    \# & $V/\Lambda$ & allowable & disallowable \\ \midrule
    \textcircled{\raisebox{-0.9pt}{1}} & $1 + \cos(\phi/f)$ & $[2, 6.09)$ & $[6.09,10]$ \\
    \textcircled{\raisebox{-0.9pt}{2}} & $1 - \phi^2/\mu^2$ & $[10^{1/2},10^4]$ & \\
    \textcircled{\raisebox{-0.9pt}{3}} & $1 - \phi^4/\mu^4$ & $[10^{-1},10^2]$ & \\
    \textcircled{\raisebox{-0.9pt}{4}} & $1 - \exp(-q\phi)$ & $[10^{-3}, 10^3]$ & \\
    \textcircled{\raisebox{-0.9pt}{5}} & $1 - \mu^2/\phi^2$ & $[10^{-6},10^3]$ & \\
    \textcircled{\raisebox{-0.9pt}{6}} & $1 + \alpha \log \phi$ & $[10^{-3},10]$ & \\
    \textcircled{\raisebox{-0.9pt}{7}} & $\left[1 - \exp \left( -\sqrt{2} \phi/\sqrt{3 \alpha} \right) \right]^2$ & $[10^{-1},93.9)$ & $[93.9,10^4]$ \\
    \textcircled{\raisebox{-0.9pt}{8}} & $\phi^p$ & $[1/2,1.05)$ & $[1.05,7/2]$ \\
    \bottomrule
\end{tabular} \label{thetable}
\end{table}

Figure~\ref{thefigure} gives a representation of these KS constraints in terms of predictions for two key observables associated with the spectrum of CMB fluctuations generated by inflation. The figure shows the values of the scalar tilt $n_s$ and the tensor-to-scalar ratio $r$ predicted by the eight different inflationary models listed in the table above. (Encircled numbers in Figure~\ref{thefigure} correspond to the number of the model in the table.) Inflationary universes that remain allowable by KS for 50--60 $e$-folds are indicated in green whereas those for which the KS criterion fails before the end of inflation are shown in red. Superposed on these theoretical predictions are the observational constraints following from the 2018 Planck TT,TE,EE+lowE+lensing analysis \cite{Planck:2018jri} ($68\:\!\%$ and $95\:\!\%$ confidence levels), indicated in purple, and, in blue, the constraints with the combined 2018 BICEP/Keck data and BAO added \cite{BICEP:2021xfz}. We see that in this set of models, which we believe to be representative, the KS criterion translates into an upper bound on the tensor-to-scalar ratio of $r \lesssim 0.08$.

\section{Discussion} \label{discussionsec} \noindent
We have given strong evidence that the semiclassical no-boundary wave function, augmented with the Kontsevich--Segal criterion, selects inflationary models with a relatively low tensor-to-scalar ratio $r \lesssim 0.08$ in the microwave background anisotropies. This upper bound on $r$ is in accordance with the current observational constraints, yet it leaves room for a future detection of gravitational waves from inflation.

Our results indicate that the KS criterion can be viewed as a refinement of the no-boundary theory of the quantum state that sharpens its predictions. In models of inflation with larger values of $r$, no-boundary saddles fail to satisfy the KS criterion when the universe becomes large. One might wonder what can possibly cause the KS criterion to fail during the quasiclassical slow-roll phase. Among the $p$-form criteria in \eqref{allowable1} it turns out that the ``$0$-form'' criterion $\operatorname{Re} \sqrt{g} > 0$ fails. This criterion, which is related to the convergence of the path integral of a massive scalar on $(M = B^4,g)$, is saturated by the extremal curve described by \eqref{extremalcurve} whereas the higher-form criteria are not. The failure of allowability during inflation can thus be attributed to the late-time development of a tachyon in the spectrum of scalar perturbations around the Hartle--Hawking solution. This being said, our analysis indicates that the KS criterion does probe the fine details of the quantum origin of inflation, for the latter are encoded precisely in the exponentially small corrections to the late-time saddle-point geometries that determine whether or not they satisfy the criterion. Indeed our results lend further credence to the raison d'\^etre of quantum cosmology, namely that a quantum gravitational completion of inflation can have verifiable observational consequences.

As an aside, we note that the alternative tunneling wave function of the universe \cite{Vilenkin1982}, constructed via a gravitational path integral \cite{Vilenkin:2018dch}, fails to meet the KS criterion. The semiclassical tunneling wave function involves gravitational instantons that belong to the fourfold degenerate family of no-boundary solutions that we discussed below \eqref{allowable3}. When evaluating their action, however, one chooses the opposite sign for $\operatorname{Re} \sqrt{g}$ along the curve $\gamma : 0 \rightarrow v$ compared to what the KS criterion demands, viz.\ $\operatorname{Re} \sqrt{g} < 0$ instead of $\operatorname{Re} \sqrt{g} > 0$. This is consonant with the observation that the ``naive'' wave function of fluctuations in the tunneling state appears to be non-normalizable (cf.~\cite{Halliwell:1989dy}). Instead it appears that a well-behaved wave function of fluctuations in the tunneling state would have to be based on a complexified integration contour for matter field fluctuations \cite{Vilenkin:2018dch}, thereby evading the KS criterion altogether.

Ultimately, the utility of quantum cosmology lies in the fact that a theory of the quantum state combined with the structure of the low-energy scalar potential yields a cosmological measure that specifies a theoretical prior for observations (see e.g.~\cite{Hartle:2013oda,Hartle:2015vfa,Hartle:2016tpo}). In this paper we have considered but the simplest quantum completion of inflation, in which an observable phase of slow-roll emerges directly from a no-boundary origin. It would be interesting to study the implications of the KS criterion in more elaborate models of initial conditions. For example, one could take a more expansive view and conceive of the range of models of inflation as different slow-roll patches in a landscape potential. In this context, the no-boundary amplitude of different backgrounds and fluctuations implies a relative weighting over different landscape regions and hence over cosmological observables that differentiate between regions. Crucially, predictions for observations follow from conditional probabilities. The ``bare'' no-boundary weighting favors backgrounds starting at a low value of the potential, followed by only a few $e$-folds of slow-roll inflation. However, no-boundary probabilities conditioned on a sufficiently accurate description of our observational situation favor slow-roll backgrounds originating on a flat plateau-like patch of the scalar potential where the conditions for eternal inflation hold \cite{Hartle2011,Hertog2014}. In future work we intend to extend our analysis into this regime and determine whether our findings are sharpened or modified in this more elaborate setting.

It would also be interesting to understand how the KS criterion relates to the swampland program. At first sight there appears to be a certain tension between both approaches, because KS appears to favor near-de Sitter saddles whereas the swampland points towards a short-lived inflationary phase. On the other hand, both considerations seem to align on a relatively low tensor-to-scalar ratio. It would be very interesting to study whether KS and the swampland are somehow two different ways of saying the same thing, or whether they are genuinely at odds with one another.

\section*{Acknowledgments} \noindent
We thank C.~de Rham, V.~Gorbenko, M.~Kleban, J.-L.~Lehners, M.~Mirbabayi, T.~Padilla, A.~Tolley and A.~Vilenkin for insightful discussions. We thank W.~P.~McBlain for his help plotting the Planck data in Figure~\ref{thefigure} and Ivan Girotto for his help setting up parallel tasking on ICTP's ARGO cluster. T.~H. and J.~K. acknowledge support from the PRODEX grant LISA - BEL (PEA 4000131558), the FWO Research Project G0H9318N and the inter-university project iBOF/21/084. J.~K. is also supported by the Research Foundation - Flanders (FWO) doctoral fellowship 1171823N.

\bibliographystyle{klebphys2}
\bibliography{refs}

\clearpage
\widetext
\phantomsection{}
\pdfbookmark[chapter]{Supplemental Material}{sm}
\begin{center}
\textbf{\large Supplemental Material}
\end{center}

\setcounter{equation}{0}
\setcounter{figure}{0}
\setcounter{table}{0}
\setcounter{page}{1}
\renewcommand{\theequation}{S\arabic{equation}}
\renewcommand{\thefigure}{S\arabic{figure}}
\renewcommand{\thetable}{S\arabic{table}}
\renewcommand{\theHequation}{equation.S\arabic{equation}}
\renewcommand{\theHfigure}{figure.S\arabic{figure}}
\renewcommand{\theHtable}{table.S\arabic{table}}

\section*{Allowable and disallowable no-boundary saddles: an illustration}\noindent
To illustrate our method of determining whether a given three-geometry specified by $(b,\chi)$ can be prepared by a no-boundary saddle that satisfies the KS criterion, consider model \textcircled{\raisebox{-0.9pt}{1}} in Table \ref{thetable}, i.e., natural inflation:
\begin{equation} \label{naturalinflationpotential}
    V(\phi) = \Lambda \left[ 1 + \cos \left( \phi / f \right) \right] \,,
\end{equation}
where $\Lambda, f > 0$ (in Planck units). By rescaling the scale factor like $a \rightarrow a/\sqrt{\Lambda}$ and the radial coordinate on the $B^4$ like $r \rightarrow r/\sqrt{\Lambda}$, $\Lambda$ is effectively set to unity in the EOM \eqref{EOM}. The metric in \eqref{NBbackground} is multiplied by an overall factor $1/\Lambda$, but this factor does not appear in the allowability condition \eqref{allowable3}. Thus for our purposes we may set $\Lambda = 1$.

We will consider a period of inflation in this model where $\phi$ evolves from near the origin towards the first minimum of the potential to the right of the origin, at $\phi = \pi f$. Since we are particularly interested in large three-geometries that are produced after a significant period of expansion (although the procedure we outline and the accompanying Mathematica code are applicable in other cases too), we will set $\chi$ to the value corresponding to the end of slow-roll inflation, which in this case is when $\varepsilon = (V'/V)^2/2 = 1$:
\begin{equation}
    \chi = 2f \, \tan^{-1} \left( \sqrt{2} f \right) .
\end{equation}
We will be interested in various sizes of the universe, corresponding to various amounts of inflation. As described in the main text, we approximate the number of $e$-folds by $\log b H(\chi)$, with an $\mathcal{O}(1)$ error. Hence, $N_e$ $e$-folds of inflation corresponds to setting
\begin{equation}
    b = \frac{e^{N_e}}{H(\chi)} \,,
\end{equation}
where $H(\chi) = \sqrt{V(\chi)/3}$.

The first step is to solve the saddle-point EOM \eqref{EOM} together with the boundary data \eqref{initSP}--\eqref{finalBD}. Specifying $N_e$ and $\varepsilon = 1$ fixes $b$ and $\chi$ through the above relations. \eqref{EOM} can then be integrated numerically along curves in the complex $r$-plane, given the no-boundary initial conditions \eqref{initSP} specified by $\phi_0 = \phi(0)$. (To avoid problems with the square in the first Friedmann--Lema\^itre equation, we use the second one instead.) Hence, we are left with the problem of finding $\phi_0$ and the endpoint $r = v$ such that \eqref{finalBD} is satisfied, given $b$ and $\chi$. We solve this numerically by integrating \eqref{EOM} along a straight line ending at $v$. This requires initial guesses for $\phi_0$ and $v$.

For $\phi_0$ an accurate estimate is
\begin{equation}
    \operatorname{Re} \phi_0 \approx \phi_i \,, \qquad\quad
    \operatorname{Im} \phi_0 \approx - \frac{\pi}{2} \frac{V'(\phi_i)}{V(\phi_i)} \,,
\end{equation}
where $\phi_i$ is the value of $\phi$ at the beginning of inflation \cite{Janssen:2020pii}. The slow-roll equations determine $\phi_i$ as a function of $N_e$ and $\chi$ via (e.g.~\cite{Janssen:2020pii})
\begin{equation}
    N_e = \int_\chi^{\phi_i}\! \di \phi \, \frac{V}{V'} \,.
\end{equation}
For $v$ we use the estimate
\begin{equation}
    \operatorname{Re} v \approx \frac{\pi}{2 H_i} \,, \qquad\quad
    \operatorname{Im} v \approx \frac{1}{H_i} \cosh^{-1} \left( b H_i \right),
\end{equation}
where $H_i = H(\phi_i)$, and parametrize the deviation from this by $\delta/H(\chi)$. This choice of variables is somewhat arbitrary, but convenient for then the solutions have $\abs{\operatorname{Re} \delta} \ll 1$, $\operatorname{Im} \delta = \mathcal{O}(1).$ This is because $\delta = 0$ would correspond to the pure dS solution with Hubble radius $1/H_i$, approximately equal to the size of the no-boundary half-sphere (e.g.~\cite{Janssen:2020pii}), and slow-roll is a small correction to this.

For definiteness let us focus on the cases $f = 4$ and $N_e \geq 5$. As an example, for $N_e = 5$ we obtain $\phi_0 \approx 7.76 \, e^{0.076 i}$, $v \approx 3.37 + 20.9 i$, while for $N_e = 25$ we obtain $\phi_0 \approx 3.65 \, e^{0.053 i}$, $v \approx 2.14 + 52.7 i$ (see the Mathematica code accompanying this supplement). Note that there are two complex boundary conditions ($a(v) = b$, $\phi(v) = \chi$) and two complex parameters ($\phi_0$, $v$) to be determined in this procedure. The numerical error in the result can be decreased by increasing the precision in all the steps of the algorithm. A well-known numerical complication is that one cannot start integrating \eqref{EOM} at exactly $r = 0$ because the boundary condition $a(0) = 0$ implies a singularity there. Instead one starts integrating at a sufficiently small value -- determined by the condition that the results are insensitive to this value -- which decreases as the precision is increased.

The next step is to determine the extremal curve $\gamma_\mathsf{e}$ by solving \eqref{extremalcurve} with the boundary condition $\lim_{\ell \rightarrow 0} \arg \gamma_\mathsf{e}(\ell) = \pi/8$. This value is independent of the parameterization of $\gamma_\mathsf{e}$ and may thus be determined by setting $\gamma_\mathsf{e} = \ell \, e^{i \theta_0}$ near $\ell = 0$. Using $a(\gamma) \approx \gamma$ around $r \approx 0$ it readily follows from \eqref{extremalcurve} that $\theta_0 = \pi/8$. To solve \eqref{extremalcurve} it would appear that the values $a(r)$ must be known for general $r \in \mathbb{C}$. With $\phi_0$ at hand, we may in principle determine these (for example, analogously as above, by integrating along a straight line connecting the origin to $\gamma$), but this procedure is computationally expensive. A much more efficient method is to simultaneously solve for $\gamma_\mathsf{e}(\ell)$ and $a(\gamma_\mathsf{e}(\ell))$ (and $\phi(\gamma_\mathsf{e}(\ell))$) by changing variables from $r$ to $\ell$ in \eqref{EOM} via $r = \gamma_\mathsf{e}(\ell)$. To do this a particular parametrization of $\gamma_\mathsf{e}$ must be chosen, for example the unit speed parameterization implied by the solution
\begin{equation} \label{extremalparam}
    \gamma_\mathsf{e}'(\ell) = i \, e^{-3 i |\arg a(\gamma_\mathsf{e}(\ell))|}
\end{equation}
to \eqref{extremalcurve}. Since $a(\gamma_\mathsf{e}(\ell)) \approx \gamma_\mathsf{e}(\ell)$ near $\ell = 0$, this determines $\gamma(\ell) \approx \ell \, e^{i \pi / 8}$ near $\ell = 0$.

\begin{figure}[p]
\centering
\includegraphics[width=86.45mm]{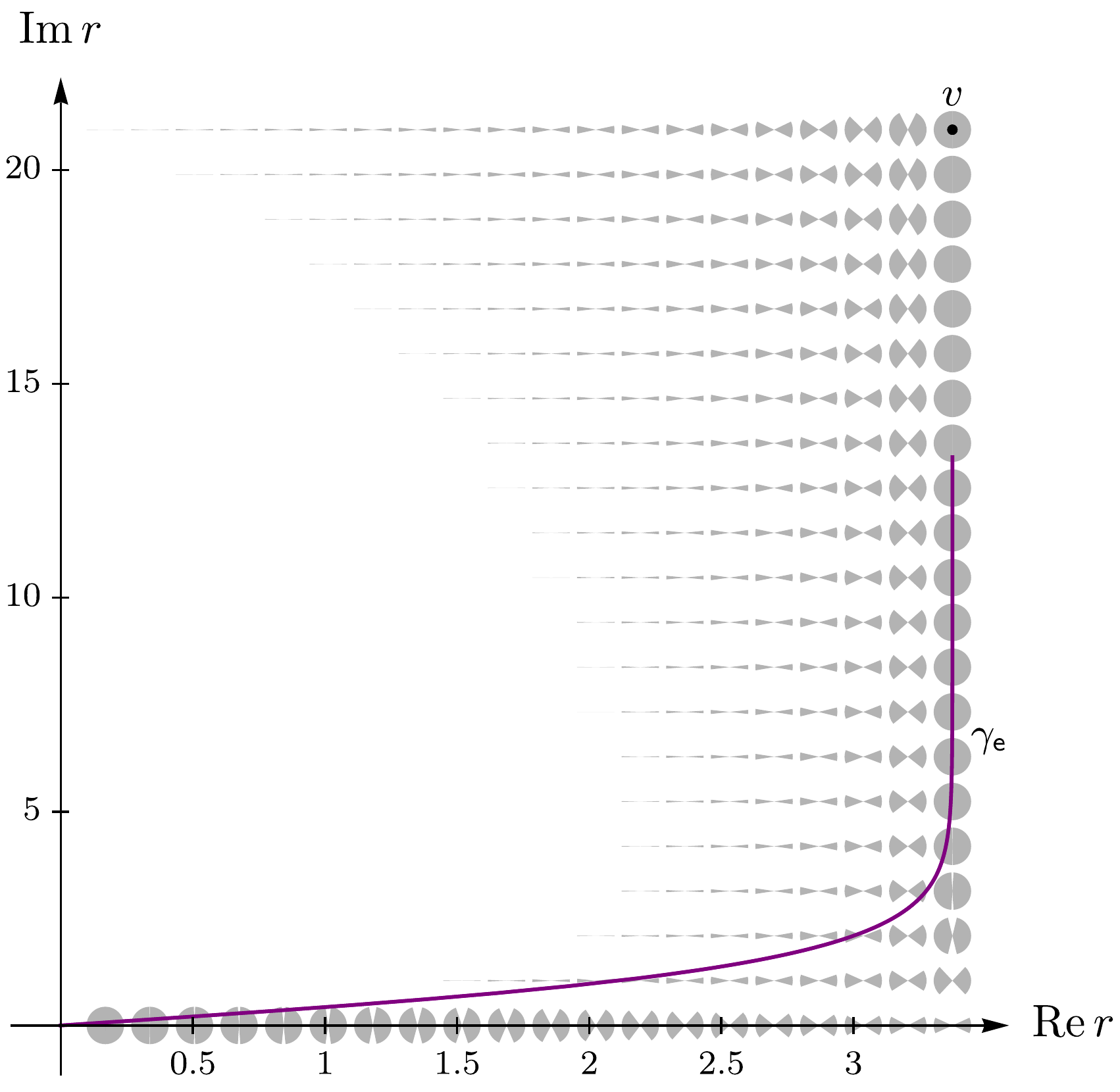}
\caption{The extremal curve $\gamma_\mathsf{e}$ (purple) for the case $f = 4$, $N_e = 5$. The cones indicate which directions $\gamma'$ satisfy the KS criterion \eqref{allowable3}, which depend on the position $\gamma \in \mathbb{C}$ because of the second term in \eqref{allowable3}. In the region without cones $3\abs{\arg{a^2}} \geq \pi$ and there are no allowable directions. Close to the origin the cones close off quickly when moving vertically, which is why $\gamma_\mathsf{e}$ has to start at an angle $\pi/8$ with the positive real axis. The question of allowability translates into the question whether $v$ is inside the ``right allowability-cone'' of the origin, the upper boundary of which is given by the extremal curve $\gamma_\mathsf{e}$. In this example, the integration was stopped when $\operatorname{Re} \gamma_\mathsf{e} = \operatorname{Re} v$ as opposed to $\operatorname{Im} \gamma_\mathsf{e} = \operatorname{Im} v$. Since all allowable curves $\gamma$ must be right-moving and lie below $\gamma_\mathsf{e}$, we conclude there can be no allowable $\gamma$ which connects the origin to the endpoint $v$ in this case.}
\label{S1thefigure}
\end{figure}

\begin{figure}[p]
\centering
\includegraphics[width=86.45mm]{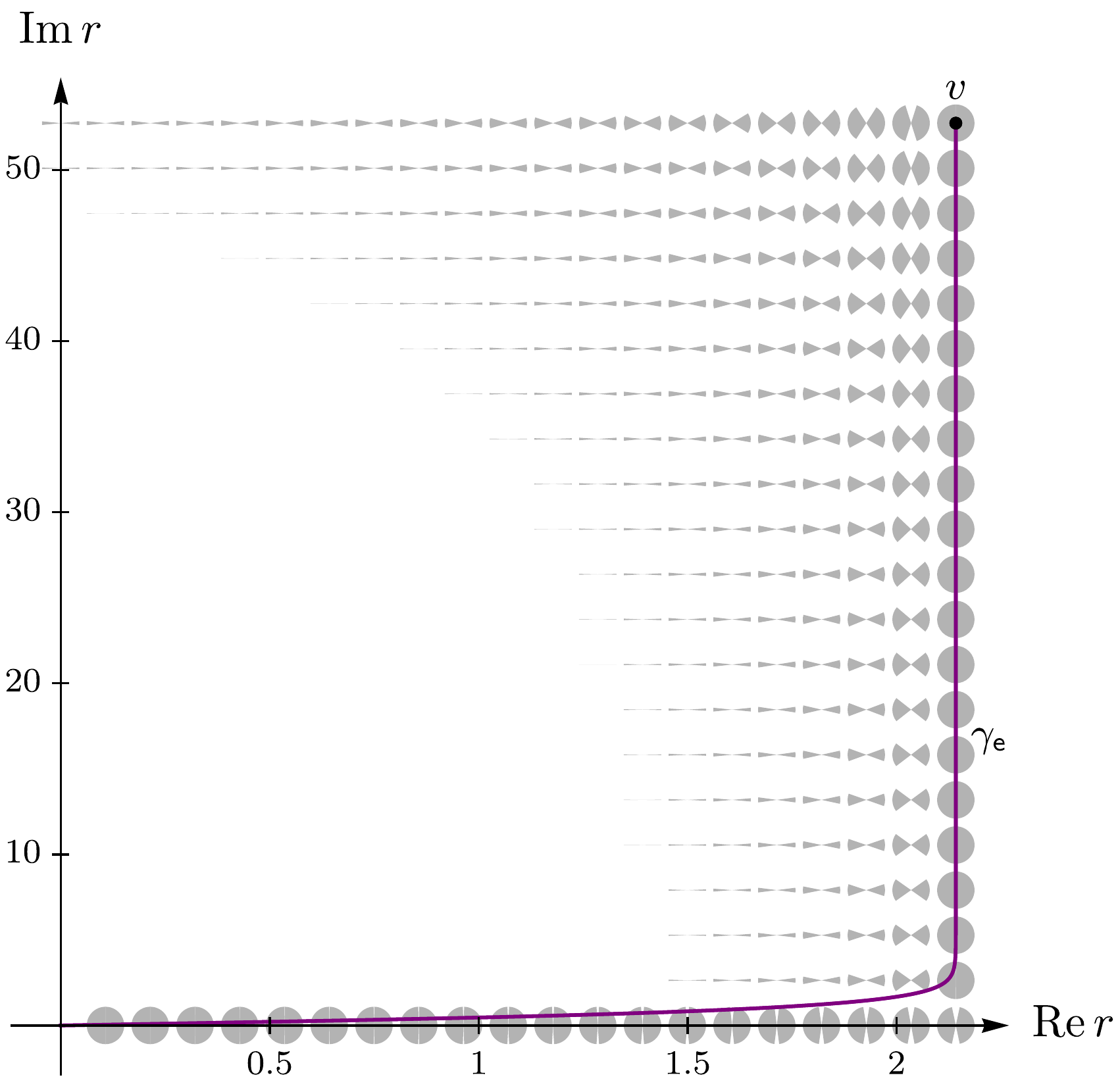}
\caption{The extremal curve $\gamma_\mathsf{e}$ (purple) for the case $f = 4$, $N_e = 25$. Here the integration was stopped when $\operatorname{Im} \gamma_\mathsf{e} = \operatorname{Im} v$ as opposed to $\operatorname{Re} \gamma_\mathsf{e} = \operatorname{Re} v$; $\gamma_\mathsf{e}$ thus stops to the \textit{left} of $v$ (not visible in the figure -- the distance between the point where $\gamma_\mathsf{e}$ was stopped and $v$ is $\mathcal{O}(10^{-36})$, see Figure~\ref{Sfig:convergence}). By making $\gamma_\mathsf{e}$ ``less extremal'', that is, by decreasing the RHS in its defining equation \eqref{extremalcurve}, we arrive at a situation which is qualitatively like the one in Figure~\ref{S1thefigure} where $\gamma_\mathsf{e}$ passes \textit{below} $v$. By continuity we conclude there must exist a curve everywhere satisfying \eqref{allowable3} and reaching the endpoint $v$.}
\label{S2thefigure}
\end{figure}

We then numerically integrate the adjusted EOM \eqref{EOM} (which are now solved along the extremal curve) and \eqref{extremalparam} with the boundary conditions
\begin{align}
    \gamma_\mathsf{e}(\ell) &= e^{i \pi / 8} \ell + \mathcal{O}(\ell^3) \,, \\
    a(\gamma_\mathsf{e}(\ell)) &= e^{i \pi/8} \ell + \mathcal{O}(\ell^3) \,, \\
    \phi(\gamma_\mathsf{e}(\ell)) &= \phi_0 + \frac{V'(\phi_0)}{8} e^{i \pi/4} \ell^2 + \mathcal{O}(\ell^4)
    \mathrlap{\quad \text{as } \ell \rightarrow 0 \,.}
\end{align}
In practice, we stop integrating when either $\operatorname{Re} \gamma_\mathsf{e} = \operatorname{Re} v$ or $\operatorname{Im} \gamma_\mathsf{e} = \operatorname{Im} v$, whichever condition is satisfied first. (In the main text we only mentioned integrating until $\operatorname{Re} \gamma_\mathsf{e} = \operatorname{Re} v$, but in practice we found the two approaches to be equivalent.) If the former condition is satisfied first, we conclude that $(b,\chi)$ cannot be prepared by an allowable no-boundary saddle with $O(4)$-symmetry on $B^4$. This is because, as we discussed in \S\ref{NBsec} of the main text, an allowable curve is always right-moving, and all curves satisfying the KS criterion must lie below the extremal one. If on the other hand the latter condition is satisfied first, we conclude that $(b,\chi)$ \textit{can} be prepared by a no-boundary solution satisfying the KS criterion. This is because if we decrease the RHS in the defining equation \eqref{extremalcurve} of the extremal curve, we generate an allowable curve lying below $\gamma_\mathsf{e}$ for which eventually the condition $\operatorname{Re} \gamma_\mathsf{e} = \operatorname{Re} v$ is satisfied first. Assuming this process is continuous, this guarantees the existence of an allowable curve reaching the desired endpoint. In Figures \ref{S1thefigure}-\ref{S2thefigure} we illustrate this for the two examples with $f = 4$, $N_e = 5$ (disallowable) and $f = 4$, $N_e = 25$ (allowable) considered previously. The configuration $(b,\chi)$ corresponding to 5 $e$-folds of natural inflation with $f = 4$ is thus suppressed in the no-boundary state enhanced by the KS criterion compared to the naive state which does not take KS into account. For 25 $e$-folds on the other hand the saddle is allowable and the amplitude is thus given by its usual ``unsuppressed'' value in the no-boundary state.

\begin{figure}[ht!]
\centering
\subfloat[]{\includegraphics[width=86.45mm]{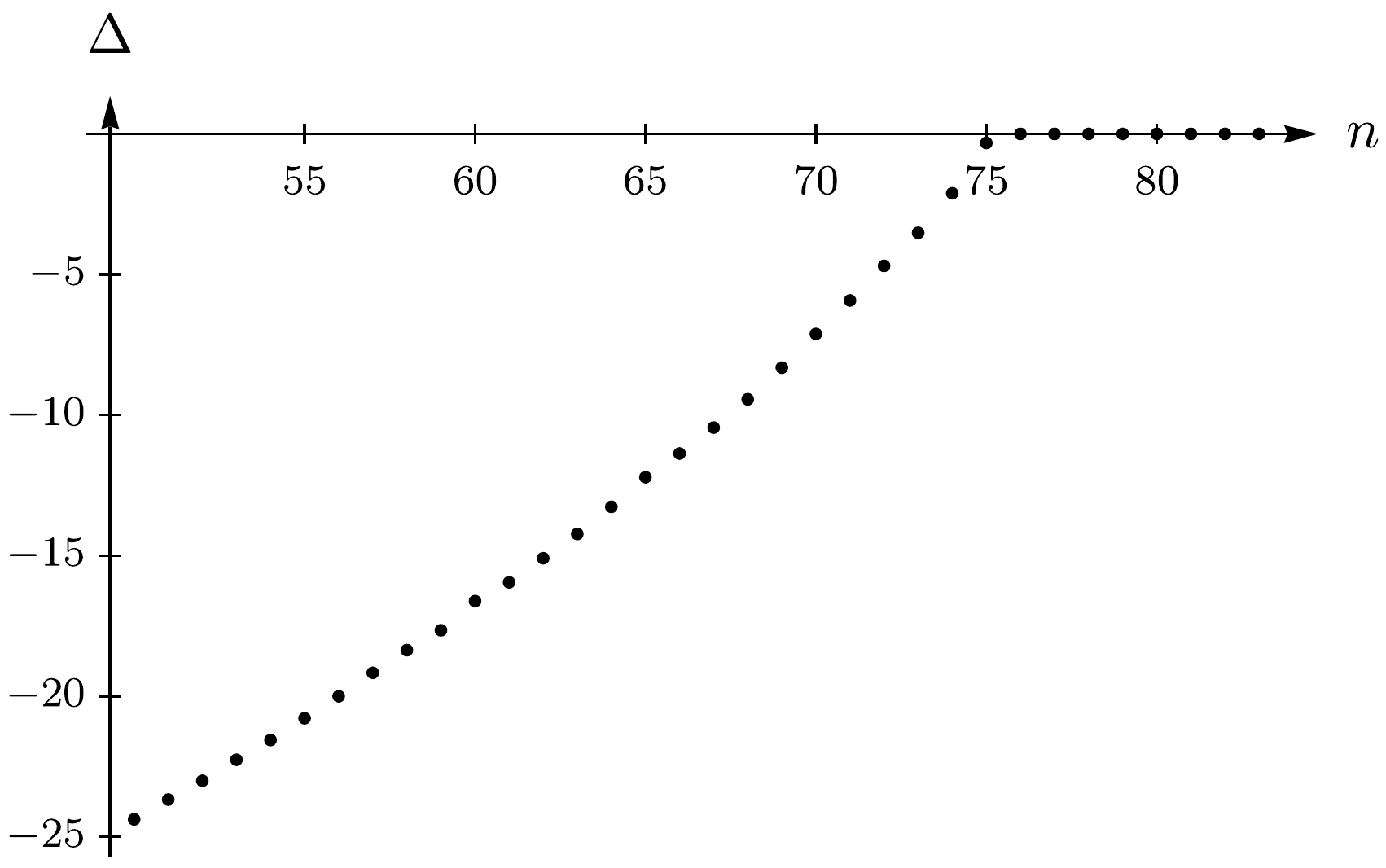}}
\hfil
\subfloat[]{\includegraphics[width=86.45mm]{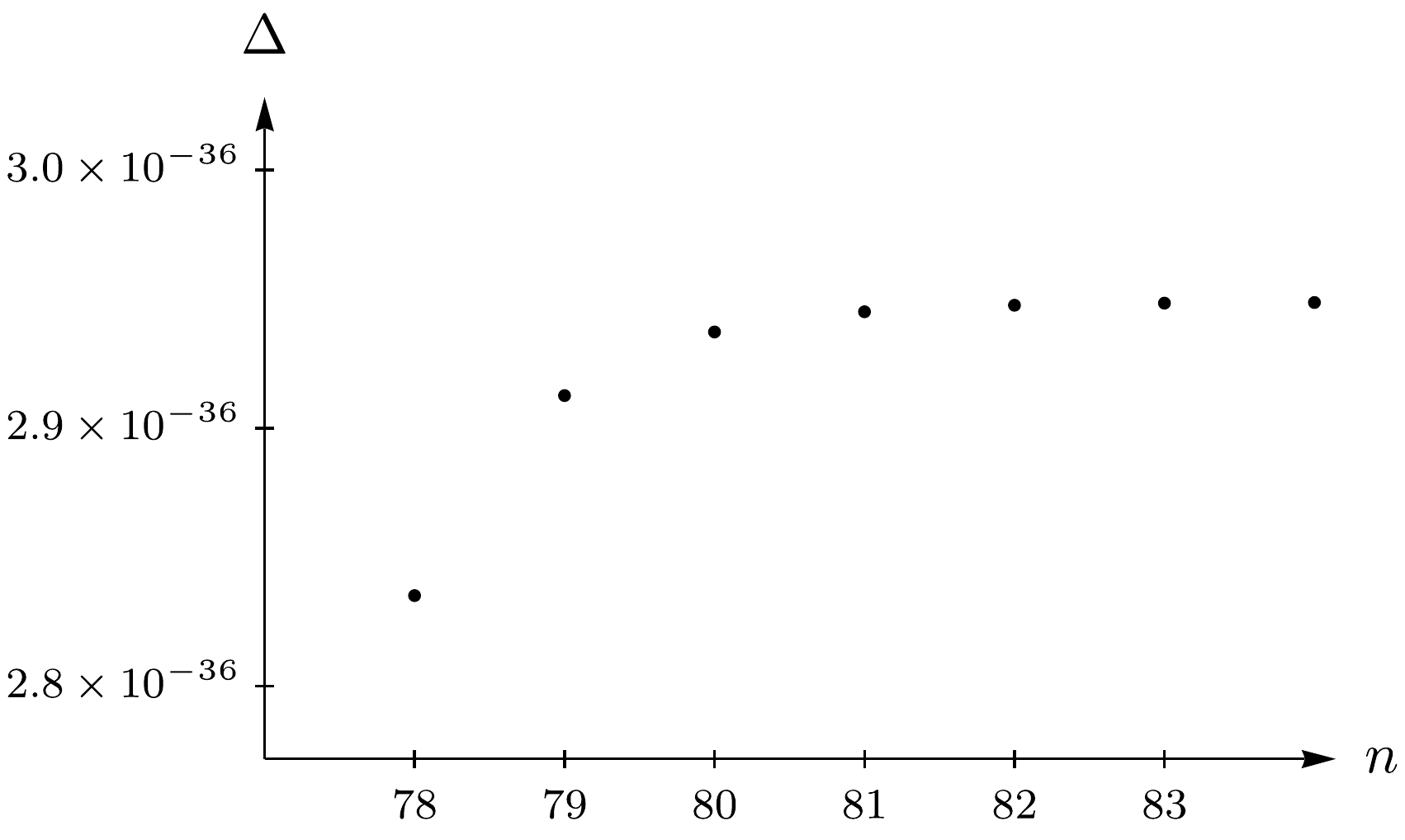}}
\caption{For the case $f = 4, N_e = 25$ we show the signed distance $\Delta$ between where the extremal curve was stopped by our numerical algorithm (when either $\operatorname{Re} \gamma_\mathsf{e} = \operatorname{Re} v$ or $\operatorname{Im} \gamma_\mathsf{e} = \operatorname{Im} v$) and the endpoint $v$, with a positive (negative) sign for horizontal (vertical) distances, versus the amount of digits $n$ kept in intermediate computations. The right figure (b) shows only the highest values of $n$ from the left figure (a) to illustrate the numerical convergence. That $\Delta$ converges to a positive value implies that the saddle is allowable by the logic explained in the text. The converged result, which is $\mathcal{O}(10^{-36})$, may seem small but this is expected since $\operatorname{Re}(\gamma_\mathsf{e} - v)$ decreases exponentially with $\operatorname{Im} \gamma_\mathsf{e}$ in the pure dS case \eqref{dSasymptotics}.}
\label{Sfig:convergence}
\end{figure}

Note that the extremal curve $\gamma_\mathsf{e}$ ends very close to the endpoint $v$ in Figure~\ref{S2thefigure}; very close compared to the distance over which $\arg a^2$ changes. This, which is also the case in the other models, confirms that $\gamma_\mathsf{e}$ can be deformed into a curve that ends at $v$ and satisfies \eqref{allowable3} strictly (verifying the above continuity assumption), whereas a priori there could have been a disallowable region where $3\abs{\arg{a^2}} \geq \pi$ separating the endpoint of $\gamma_\mathsf{e}$ from $v$, rendering such a deformation impossible. In Figure~\ref{Sfig:convergence} we show in the $N_e = 25$ case how we verified that the numerics have converged. We performed analogous convergence checks for all the data going into Figure~\ref{thefigure} (see supplemental data files).

As a concluding remark, notice that in this model (potential \eqref{naturalinflationpotential} with $f = 4$) the boundary conditions $(b,\chi)$ go from being disallowable to allowable as we increase the amount of $e$-folds of expansion. It turns out the transition happens at around $N_e = 20.8$ $e$-folds of inflation in this case. This feature is not general: in the monomial potential $V(\phi) = \phi^{6/5}$ the boundary conditions go from being allowable at small values of $N_e$ to disallowable at values larger than about $N_e = 6.15$. From the cases we studied it would appear that concave potentials (at $\phi_0$) follow the former rule while convex potentials follow the latter. Our main result in Figure~\ref{thefigure} is consistent with this rule of thumb.

\newpage
\section*{Mathematica code and data files}\noindent
The accompanying Mathematica code illustrates how to obtain the results of the detailed examples in the previous section, including how to find the saddles, determine if a saddle is allowable, check numerical convergence, find the transition between allowable and disallowable and produce Figures~\ref{S1thefigure}--\ref{Sfig:convergence}, and how to reproduce results from the data files. The data files contain the numerical results that went into producing Figure~\ref{thefigure} and Table~\ref{thetable} in the main text. The data is structured as \texttt{csv} tables with the first line of each file describing the content of that column; see Table~\ref{Stab:symbols} for an explanation of the symbols.

\begin{table}[h!]
\centering
\caption{Description of data file header symbols. Note that $\abs{\phi_0}$, $\gamma = \arg \phi_0$, $\operatorname{Re} v$ and $\operatorname{Im} v$ are the four real parameters specifying the saddle. Recall $r=v$ is the point where $a(v) = b$ and $\phi(v) = \chi$.}
\begin{tabular}{ll}
    \toprule
    symbol & description \\ \midrule
    \texttt{wp} & number of digits of precision \\
    \texttt{logbH} & number of $e$-folds $\log b H(\chi)$ \\
    \texttt{fd} & parameter in potential, see Table~\ref{thetable} in the main text \\
    \texttt{fdCrit} & parameter in potential at transition between allowable and disallowable \\
    \texttt{modphi0} & $\abs{\phi_0}$, absolute value of the scalar field at $r = 0$ \\
    \texttt{gamma} & $\gamma = \arg{\phi_0}$, phase of the scalar field at $r = 0$ \\
    \texttt{vX} & $\operatorname{Re} v$, real part of the endpoint $v \in \mathbb{C}$ \\
    \texttt{vY} & $\operatorname{Im} v$, imaginary part of $v$ \\
    \texttt{Delta} & $\Delta$, signed distance between the endpoint of the extremal curve and \\
    & $v$, with a positive (negative) sign for horizontal (vertical) distances \\
    \bottomrule
\end{tabular} \label{Stab:symbols}
\end{table}

The names of the data files are of the format \texttt{<model>\_[<range>]\_<type>.csv} where \texttt{<model>} is a number from Table~\ref{thetable} in the main text specifying the potential, \texttt{<range>} specifies the parameter range according to Table~\ref{Stab:ranges} (when applicable) and \texttt{<type>} is one of \texttt{Delta}, \texttt{Delta\_fdCrit} and \texttt{fdCrit}.

The reason behind the \texttt{<range>} parameter is that for some potentials it depends on the value of the parameter in the potential whether the condition $\varepsilon = 1$ or $\abs{\eta} = 1$, defining the end of slow-roll, is reached first when rolling down. Note that the exact criterion is somewhat arbitrary and for some models the \texttt{small} and \texttt{large} data files have slightly overlapping ranges of the parameter in the potential. In model \textcircled{\raisebox{-0.9pt}{6}}, we actually investigated the model for $10^{-3} \leq \alpha \leq 10$ (see Table~\ref{thetable} in the main text) even though $\abs{\eta} = 1$ only is the correct criterion for the end of slow-roll up to $\alpha \approx 6.52$ (see Table~\ref{Stab:ranges}). Similarly in model \textcircled{\raisebox{-0.9pt}{8}} the criterion $\varepsilon = 1$ is only correct down to $p = 2/3$ but we use it all the way to $p = 1/2$. Since none of these regions are close to a transition between allowable and disallowable this does not matter for the main result.

Files with $\texttt{<type>}=\texttt{Delta}$ contain a scan over parameters to identify if there are any disallowable saddles at all in that parameter range of the given model, $\texttt{<type>}=\texttt{fdCrit}$ contain critical potential parameters at the border between allowability and disallowability found by searching for roots of $\Delta$, defined in Table~\ref{Stab:symbols}, and $\texttt{<type>}=\texttt{Delta\_fdCrit}$ contain evaluations of $\Delta$ used to locate the critical parameters. When finding the transition we use $\texttt{AccuracyGoal}\to 5$ and $\texttt{PrecisionGoal}\to 5$ as tolerances for the position of the root and no tolerance criterion for $\abs{\Delta}$ to limit excessive computations (see supplemental Mathematica code).

\begin{table}[h!]
\centering
\caption{Investigated parameter ranges for the parameters in the potentials in Table~\ref{thetable} of the main text, including the \texttt{<range>} part of the data filenames (when applicable). The last column of this table specifies which criterion was used to define the end of slow-roll in the corresponding data files while the third column provides the actual transition point between the criteria.}
\begin{tabular}{clll}
    \toprule
    \# & \texttt{<range>} & parameter range & end of slow-roll \\ \midrule
    \textcircled{\raisebox{-0.9pt}{1}} & n/a & $f \in [2, 10]$ & $\varepsilon = 1$ \\
    \textcircled{\raisebox{-0.9pt}{2}} & n/a & $\mu \in [10^{1/2}, 10^4]$ & $\varepsilon = 1$ \\
    \textcircled{\raisebox{-0.9pt}{3}} & \texttt{small} & $\mu \in [10^{-1}, 540^{1/4}]$ & $\abs{\eta} = 1$ \\
    \textcircled{\raisebox{-0.9pt}{3}} & \texttt{large} & $\mu \in [540^{1/4}, 10^2]$ & $\varepsilon = 1$ \\
    \textcircled{\raisebox{-0.9pt}{4}} & \texttt{small} & $q \in [10^{-3}, 2^{-1/2}]$ & $\varepsilon = 1$ \\
    \textcircled{\raisebox{-0.9pt}{4}} & \texttt{large} & $q \in [2^{-1/2}, 10^3]$ & $\abs{\eta} = 1$ \\
    \textcircled{\raisebox{-0.9pt}{5}} & \texttt{small} & $\mu \in [10^{-6}, 3(3/2)^{1/2}]$ & $\abs{\eta} = 1$ \\
    \textcircled{\raisebox{-0.9pt}{5}} & \texttt{large} & $\mu \in [3(3/2)^{1/2}, 10^3]$ & $\varepsilon = 1$ \\
    \textcircled{\raisebox{-0.9pt}{6}} & n/a & $\alpha \in [10^{-3}, 6.52]$ & $\abs{\eta} = 1$ \\
    \textcircled{\raisebox{-0.9pt}{7}} & n/a & $\alpha \in [10^{-1}, 10^4]$ & $\varepsilon = 1$ \\
    \textcircled{\raisebox{-0.9pt}{8}} & \texttt{small} & $p \in [2/3, 2]$ & $\varepsilon = 1$ \\
    \textcircled{\raisebox{-0.9pt}{8}} & \texttt{large} & $p \in [2, 7/2]$ & $\abs{\eta} = 1$ \\
    \bottomrule
\end{tabular} \label{Stab:ranges}
\end{table}

\end{document}